\documentclass[twocolumn,amsmath,amssymb,prd]{revtex4-2}
\def\al{\alpha}

\def\de{\delta}
\def\ep{\epsilon}

\def\th{\theta}

\def\la{\lambda}

\def\si{\sigma}

\def\ph{\phi}

\def\om{\omega}

\def\fr#1#2{{{#1}\over{#2}}}
\def\frac#1#2{{\textstyle{{#1}\over{#2}}}}

\def\ol{\overline}

\def\lsim{\mathrel{\rlap{\lower4pt\hbox{\hskip1pt$\sim$}}
    \raise1pt\hbox{$<$}}}
\def\gsim{\mathrel{\rlap{\lower4pt\hbox{\hskip1pt$\sim$}}
    \raise1pt\hbox{$>$}}}

\def\etal{{\it et al.}}

\def\vev#1{\langle {#1}\rangle}

\def\ket#1{|{#1}\rangle}

\def\sqr#1#2{{\vcenter{\vbox{\hrule height.#2pt
         \hbox{\vrule width.#2pt height#1pt \kern#1pt
         \vrule width.#2pt}
         \hrule height.#2pt}}}}

\newcommand{\beq}{\begin{equation}}
\newcommand{\eeq}{\end{equation}}
\newcommand{\bea}{\begin{eqnarray}}
\newcommand{\eea}{\end{eqnarray}}
\newcommand{\rf}[1]{(\ref{#1})}

\newcommand{\bM}{\begin{pmatrix}}
\newcommand{\eM}{\end{pmatrix}}

\def\nn{\nonumber}

\def\f{w}

\def\mbf#1{\boldsymbol #1}

\def\syjm#1#2{{}_{#1}Y_{#2}}

\def\V{\mathcal V}

\def\T{\mathcal T}

\def\K{\mathcal K}

\def\pvec{\mbf p}

\def\sivec{\mbf\si}

\def\pmag{|\pvec|}

\def\punit{\hat p}

\def\epunit{\hat\ep}
\def\thunit{\hat\th}
\def\phunit{\hat\ph}

\def\phat{\mbf\punit}

\def\ephat{\mbf\epunit}
\def\thhat{\mbf\thunit}
\def\phhat{\mbf\phunit}

\def\nr{{\rm NR}}

\def\nrtemplate#1#2#3{#1^{\nr#3}_{#2}}

\def\cs133{\rm Cs}

\def\Vnrf#1#2{\nrtemplate{{\V_{#1}}}{#2}{}}
\def\TzBnrf#1#2{\nrtemplate{{\T_{#1}}}{#2}{(0B)}}

\def\ToBnrf#1#2{\nrtemplate{{\T_{#1}}}{#2}{(1B)}}

\def\anrf#1#2{\nrtemplate{{a_{#1}}}{#2}{}}
\def\cnrf#1#2{\nrtemplate{{c_{#1}}}{#2}{}}

\def\sVnrf#1#2{\nrtemplate{{\V_{#1}}}{#2}{,{\rm Sun}}}
\def\sanrf#1#2{\nrtemplate{{a_{#1}}}{#2}{,{\rm Sun}}}
\def\scnrf#1#2{\nrtemplate{{c_{#1}}}{#2}{,{\rm Sun}}}
\def\sgzBnrf#1#2{\nrtemplate{{g_{#1}}}{#2}{(0B), {\rm Sun}}}
\def\sgoBnrf#1#2{\nrtemplate{{g_{#1}}}{#2}{(1B),{\rm Sun}}}
\def\sHzBnrf#1#2{\nrtemplate{{H_{#1}}}{#2}{(0B),{\rm Sun}}}
\def\sHoBnrf#1#2{\nrtemplate{{H_{#1}}}{#2}{(1B),\rm Sun}}
\def\sTzBnrf#1#2{\nrtemplate{{\T_{#1}}}{#2}{(0B),{\rm Sun}}}
\def\sToBnrf#1#2{\nrtemplate{{\T_{#1}}}{#2}{(1B),\rm Sun}}

\def\widecheck#1{\hskip#1pt\huge$\check{}$}
\def\bighacek#1#2{\vbox{\ialign{##\crcr\widecheck#2\crcr
  \noalign{\kern-9.5pt\nointerlineskip}
   $\hfil\displaystyle{#1}\hfil$\crcr}}}

\def\mr{\ol{m}_{\rm r}}

\def\k{k}

\def\chM{\vartheta}

\def\fb{{\overline{\f}}}

\def \j{j}

\def\TL{T_L}

\begin{document}

\title{Bounds on Lorentz and CPT violation from the $1S$-$2P$ transition in antihydrogen } 

\author{ Arnaldo J.\ Vargas$^1$}

\affiliation{
$^1$Laboratory of Theoretical Physics, Department of Physics, University of Puerto Rico, R\'io Piedras, Puerto Rico 00936\\
}

\begin{abstract}
A model for the Lorentz- and CPT-violating frequency shift for the antihydrogen $1S$-$2P$ transition in the presence of an external magnetic field is derived. Using the recent measurement of the $1S$-$2P$ transition frequency in antihydrogen by the ALPHA collaboration, which they demonstrated agrees with predictions from the Standard Model of particle physics, we establish the first constraints on 26 effective coefficients for Lorentz and CPT violation. Also, this work uses these results to underscore the value of measuring multiple transition frequencies to test CPT symmetry through antihydrogen spectroscopy, emphasizing the advantages of transitions involving states with higher angular momentum.
\end{abstract}

\maketitle

\section{Introduction} 

The Standard-Model Extension (SME) provides a comprehensive framework for systematically testing Lorentz symmetry \cite{ck}. This framework is motivated by the idea that tiny deviations from Lorentz symmetry could serve as low-energy signals for physics beyond the Standard Model and General Relativity \cite{ksp}. The framework associates each Lorentz-violating operator with a specific coefficient, known as a coefficient for Lorentz violation or SME coefficient. If Lorentz symmetry holds, all these coefficients must be zero. However, any non-zero value would indicate a violation of Lorentz symmetry. The SME's approach involves identifying experiments sensitive to different coefficients for Lorentz violation, allowing for a systematic search of exceptions to Lorentz symmetry. Moreover, by quantifying each test based on its sensitivity to these coefficients, the SME framework helps evaluate how competitive, complementary, or overlapping various Lorentz symmetry tests are.

Another symmetry closely related to Lorentz symmetry is CPT symmetry. A well-known result is that CPT violation implies Lorentz violation in any realistic effective field theory \cite{owg}. Consequently, the SME is also a framework for the systematic study of CPT symmetry. Some of the earliest models developed within the SME framework were for antimatter tests, including models for antihydrogen spectroscopy \cite{anH} and antiparticles in Penning traps \cite{pen}. Later, the SME was expanded to include Lorentz-violating terms affecting the motion of matter in curved spacetime \cite{ko04}, leading to models that explored the potential anomalous free-fall motion of antimatter \cite{grav}.

In its early development, the SME focused on Lorentz-violating operators with mass dimensions $d \leq 4$, known as minimal operators, to avoid introducing nonrenormalizable operators to the framework \cite{ck}. However, it soon became evident that this restriction was unnecessary and possibly counterproductive. The rationale for including nonminimal terms, operators with $d\geq 5$, in the SME follows from the idea that any underlying fundamental theory that breaks Lorentz symmetry would manifest as an effective field theory at low energies, potentially containing nonminimal operators as it does not need to be renormalizable. These nonminimal operators have a natural suppression by the ratio between the current energy scale accessible by experiments and the energy scale at which the low-energy effective theory breaks down, providing an attractive explanation for the lack of evidence of Lorentz violation. In contrast, minimal operators do not possess this inherent suppression, requiring justifying the small size of their associated coefficients by pointing to the empirical fact that no evidence for Lorentz violation currently exists. Thus, it is reasonable to expect that any Lorentz violation is more likely to be controlled by nonminimal operators rather than minimal ones.

The SME has routinely expanded its framework to include more nonminimal operators \cite{km09, km12, km13, nonmingrav, kl21}. A current focus of research is to develop models for testing Lorentz and CPT symmetry that incorporate the nonminimal operators, as they can alter the characteristics of the signals for Lorentz violation observable in different types of experiments. Models incorporating nonminimal operators already exist for antimatter experiments, such as those with antiparticles trapped in Penning traps \cite{dk16}, antihydrogen spectroscopy experiments \cite{kv15}, and antimatter gravity tests \cite{kl21}.

Significant advancements in antimatter research have been achieved in recent years, notably in high-precision measurements of antiproton properties in a Penning trap \cite{BASE} and antihydrogen transition frequencies, including the one for the ground-state splitting \cite{ah17}, the $1S$-$2S$ \cite{ah18}, and $1S$-$2P$ transitions \cite{ah18a, 1s2p}. Another recent antihydrogen measurement was its free-fall acceleration \cite{an23}. Future experiments that could improve these measurements include alternative approaches to measuring antihydrogen’s free-fall acceleration \cite{gbar, aegis} and in-beam antihydrogen spectroscopy \cite{ASACUSA}.

The first direct constraints on Lorentz- and CPT-violating coefficients from antihydrogen spectroscopy \cite{kv18} were obtained from the measured $1S$-$2S$ transition frequency by the Antihydrogen
Laser Physics Apparatus (ALPHA) collaboration \cite{ah18}. The prospects for constraining SME coefficients using transitions within the ground state of antihydrogen in the presence of a magnetic field are promising \cite{kv15}. Unfortunately, the dominant contributions to these transition frequencies due to Lorentz and CPT violation, as demonstrated in \cite{kv15}, have a structure similar to the linear-order contributions from the magnetic field. For this reason, the methods implemented in the antihydrogen experiment \cite{ah17} to minimize the contributions from the magnetic field inhomogeneity also eliminate all the contributions from Lorentz- and CPT-violating operators coming from the model developed in \cite{kv15}. The previous statement does not mean that the final published results are insensitive to Lorentz and CPT violation, as models that include higher-order terms are likely to contribute to the frequency shift. However, based on current models, imposing bounds on Lorentz and CPT violation from measurements of the $1S$ splitting will require an effort to measure the magnetically sensitive transitions as precisely as possible.

In this work, we consider the $1S$-$2P$ transition measured by the ALPHA collaboration, the third type of transition observed in antihydrogen spectroscopy \cite{1s2p}, to impose bounds on SME coefficients. Although measurements of the $1S$-$2S$ transition \cite{ah18} and the $1S$ splitting \cite{ah17} have higher precision and lower absolute uncertainty than for the $1S$-$2P$ transition \cite{1s2p}, models for Lorentz and CPT symmetry \cite{kv15} indicate that, contrary to appearance, the former transitions are not necessarily more sensitive to Lorentz and CPT violation than the latter. While the $1S$-$2P$ transition is generally less sensitive to individual SME coefficients, it receives contributions from a larger set of SME coefficients than the other transitions \cite{kv15}. Consequently, the $1S$-$2P$ transition can detect forms of Lorentz and CPT symmetry breaking that are inaccessible through measurements of the other transitions.

This work is organized as follows. Including the introduction and the summary, it contains five sections. The section following the introduction presents the derivation of the Lorentz- and CPT-violating frequency shift in a laboratory frame at the surface of the Earth. The subsequent section discusses the transformation of the frequency shift expression from the laboratory frame to the Sun-centered frame and provides an expression for the constant contribution to the frequency shift in terms of the Sun-centered frame coefficients for Lorentz violation. The penultimate section details the constraints on SME coefficients obtained by comparing these results with the recent measurement of the $1S$-$2P$ transition in antihydrogen, using the model for testing Lorentz and CPT symmetry derived in the previous sections. Throughout this work, we use a unit system where $\hbar=c=1$.

\section{Lorentz-violating frequency shift in the laboratory frame}
\label{seclab}
In this section, we derive an expression for the Lorentz-violating frequency shift for transitions $1S_{d}-2P_{c-}$ and $1S_{d}-2P_{f-}$ in the presence of a 1T magnetic field, as studied by the ALPHA collaboration \cite{1s2p}. The notation used to specify the magnetically sensitive energy states in these transitions follows the convention presented in \cite{1s2p}. The section is organized as follows. First, we define the quantum states representing the energy levels involved in the transitions. Then, we introduce the Lorentz-violating perturbation for hydrogen and antihydrogen considered in this work. Finally, we derive the Lorentz-violating energy shifts for the levels involved in both transitions and use them to calculate the resulting frequency shift.

The basis used to describe the system’s energy states is $\ket{nLm_L}\ket{m_S}\ket{m_I}$. Here, $\ket{nLm_L}$ represents the component associated with the spatial degrees of freedom for the nonrelativistic energy states of antihydrogen, where $n$ is the principal quantum number, $L$ is the orbital angular momentum quantum number, and $m_L$ corresponds to the projection of the orbital angular momentum along the direction of the magnetic field. Additionally, $\ket{m_S}$ and $\ket{m_I}$ represent the spin states of the positron and antiproton, respectively.

The state $\ket{1S_d}$, as defined in \cite{1s2p}, corresponds to the $1S$ state with $m_F = -1$, where $m_F = m_L + m_S + m_I$. It is given by
\beq
\ket{1S_{d}}=\ket{1\,0\,0}\,\ket{-}\,\ket{-},
\label{1Sstate}
\eeq
where $\ket{-}$ and $\ket{+}$ denote spin-down and spin-up states, respectively.

The states $\ket{2P_{c-}}$ and $\ket{2P_{f-}}$ are $2P$ states with $m_F = 0$ and $m_I = -1/2$, as described in \cite{1s2p}. These states are superpositions of a state with $(m_L, m_S, m_I)$ equal to $(0, 1/2, -1/2)$ and a state with $(1, -1/2, -1/2)$. They can be expressed as
\bea
\ket{2P_{c-}}&=& \lambda_1 \,\ket{2\,1\,0}\,\ket{+}\,\ket{-}+\lambda_2\,\ket{2\,1\,1}\ket{-}\ket{-},\nn\\
\ket{2P_{f-}}&=& \lambda_2 \,\ket{2\,1\,0}\,\ket{+}\,\ket{-}-\lambda_1\,\ket{2\,1\,1}\ket{-}\ket{-}.
\label{2Pstate}
\eea
The coefficients $\lambda_1$ and $\lambda_2$ depend on the strength of the applied magnetic field. When the Zeeman effect is weak compared to the fine structure but strong relative to the hyperfine structure, the coefficients are approximately $\lambda_1 = \sqrt{2/3}$ and $\lambda_2 = \sqrt{1/3}$. In this regime, the energy states are eigenstates of the positron's total angular momentum $J$. Specifically, $\ket{2P_{c-}}$ corresponds to the quantum numbers $(J, m_J) = (3/2, 1/2)$, while $\ket{2P_{f-}}$ corresponds to $(1/2, 1/2)$. In contrast, when the Zeeman effect is strong relative to the fine structure, the coefficients are approximately $\lambda_1 = 0$ and $\lambda_2 = 1$. For a magnetic field of 1T, relevant to this study, the coefficients are approximately $\lambda_1 = 0.37$ and $\lambda_2 = 0.93$.

The matrix elements of the Lorentz-violating perturbation in the unperturbed energy basis determine the Lorentz-violating energy shift. Therefore, the next natural step is introducing the perturbation to the antihydrogen considered in this study. We will start by presenting the perturbation for hydrogen, from which we derive the corresponding perturbation for antihydrogen.

The Lorentz-violating perturbation applied to the hydrogen atom in this work is described in Section II of \cite{kv15} and based on a perturbation derived in \cite{km13}. It is given by
\beq
\de h_H=\de h_{e}^\nr+\de h_{p}^\nr,
\label{hexpa}
\eeq
where $\delta h_w^\nr$ is the single-particle Lorentz-violating perturbation for the electron ($w=e$) and the proton ($w=p$). The specific form of the perturbation is given by
\bea
\de h_\f^\nr
&=& -\sum_{k \j m} \pmag^k
\syjm{0}{\j m}(\phat)
\left(\Vnrf{\f}{\k \j m}+\sivec\cdot\ephat_r\TzBnrf{\f}{k \j m}\right)\nn\\
&&+\sum_{k \j m}\pmag^k \syjm{+ 1}{\j m}(\phat) \sivec\cdot\ephat_-\ToBnrf{\f}{k \j m} \nn\\
&&-\sum_{k \j m}\pmag^k \syjm{- 1}{\j m}(\phat) \sivec\cdot\ephat_+
\ToBnrf{\f}{k \j m}.\nn\\
\label{nr}
\eea
Here, the summation index $k$ is restricted to the values $0$, $2$, and $4$. The index $j$ spans from 0 to 5, while the index $m$ ranges over the interval $-j \leq m \leq j$.

The vector $\sivec=(\si^1,\si^2,\si^3)$ represents the Pauli vector composed of the Pauli matrices $\si^i$. The definition of the unit vectors in the expression are $\ephat_r = \phat$ and $\ephat_\pm = (\thhat \pm i\phhat)/\sqrt{2}$, where the unit vectors $\thhat$ and $\phhat$ are the ones associated with the polar angle $\theta$ and azimuthal angle $\phi$ in momentum space. Specifically, these angles satisfy the relation $\phat = (\sin\th\cos\ph,\sin\th\sin\ph,\cos\th)$.  

The coefficients $\Vnrf{\f}{k \j m}$ and ${\T_\f}^{\nr(qB)}_{k\j m}$, with $q$ taking values of $0$ or $1$, represent the nonrelativistic (NR) spherical coefficients for Lorentz violation, as introduced in \cite{km13}. These coefficients are linear combinations of SME coefficients linked to operators with a definite CPT handedness. These linear combinations are given by
\bea
\Vnrf{\f}{k\j m} &=&
\cnrf{\f}{k\j m} - \anrf{\f}{k\j m},
\nonumber \\
{\T_\f}^{\nr(qP)}_{k\j m} &=&
{g_\f}^{\nr(qP)}_{k\j m} - {H_\f}^{\nr(qP)}_{k\j m},
\label{cpt}
\eea
where $a$- and $g$-type coefficients are associated with CPT-odd operators, while the $c$- and $H$-type coefficients are related to CPT-even ones. 

Each of these coefficients identifies a Lorentz-violating spherical tensor operator, in the perturbation, expressed in terms of spin-weighted spherical harmonics $\syjm{s}{j m}(\phat)$ with spin weight $s$. In this work, we use the definition of the spin-weighted harmonics provided in Appendix A of \cite{km09}. It is important to note that the conventional spherical harmonics  $Y_{j m}(\theta,\phi)=\syjm{0}{j m}(\theta, \phi)$ are recovered when $s=0$. 

The indices of the nonrelativistic coefficients specify characteristics of the corresponding Lorentz-violating operator, including the rank $j$ of the spherical tensor, the component $m$ of the tensor, and the power of the momentum $k$. Additionally, the indices $0B$ and $1B$ provide insights into the spin weight of the operator and its behavior under a parity transformation. Table IV of \cite{km13} outlines the permissible ranges of values for the indices $k$, $\j$, and $m$. The reader is advised that in this study, we follow the convention set forth in \cite{kv15}, utilizing the subscript index $k$ instead of $n$, in contrast to the convention in \cite{km13}.

The single-particle perturbation for the antiparticle, denoted as $\delta h_{\fb}$, can be derived directly from the SME Lagrange density, similar to the particle case. In this work, we define the antiparticle perturbation as the CPT-transformed counterpart to the particle one. The antiparticle perturbation $\delta h_{\fb}$ takes the same form as $\delta h_{\f}$ in \rf{nr}. The expressions for the $\V$- and $\T$-type antiparticle nonrelativistic coefficients in terms of the particle coefficients with definite CPT-handedness are
\bea
\Vnrf{\fb}{k\j m} &=&
\cnrf{\f}{k\j m} + \anrf{\f}{k\j m},
\nonumber \\
{\T_\fb}^{\nr(qP)}_{k\j m} &=&
-{g_\f}^{\nr(qP)}_{k\j m} - {H_\f}^{\nr(qP)}_{k\j m}.
\label{cptb}
\eea 
The distinction between \rf{cpt} and \rf{cptb} lies in the sign of the coefficients associated with CPT-odd operators. This sign difference results from the transformation of CPT-odd operators during the CPT transformation.

The antihydrogen Lorentz-violating perturbation is given by
\beq
\de h_{\bar{H}}=\de h_{\bar{e}}+\de h_{\bar{p}},
\label{hbar}
\eeq 
where $\de h_{\bar{e}}$ is the single-particle perturbation for the positron and $\de h_{\bar{p}}$ for the antiproton. 

After introducing the Lorentz-violating perturbation for antihydrogen, we can derive the Lorentz-violating energy shift for the energy levels involved in the transition by calculating the expectation values of the perturbation with respect to the unperturbed states in \eqref{1Sstate} and \eqref{2Pstate}.

The Lorentz-violating energy shift for the $1S_{d}$ state is 
\bea
\de \ep_{1S_d}&=&\fr{1}{\sqrt{12\pi}}\sum_{\fb,k}(\TzBnrf{\fb}{k10}+2\ToBnrf{\fb}{k10})\vev{\pmag^k}_{10}\nn\\
&&-\fr{1}{\sqrt{4\pi}}\sum_{\fb,k}\Vnrf{\fb}{k00}\vev{\pmag^k}_{10},
\label{e1S}
\eea 
where $\fb=\bar{e}$  for the positron and $\fb=\bar{p}$ for the antiproton.  The expecation values $\vev{\pmag^k}_{nL}$ are determined by
\bea
\vev{\pmag^0}_{nL}&=& 1,
\nonumber\\
\vev{\pmag^2}_{nL}&=& \left(\dfrac{\al \mr}{n}\right)^2,
\nonumber\\
\vev{\pmag^4}_{nL}&=&
\left(\dfrac{\al \mr}{n}\right)^4
\left(\dfrac{8n}{2L+1}-3\right).
\label{radialExp}
\eea
Here, $\alpha$ refers to the fine-structure constant, $\bar{m}_r$ denotes the reduced mass for antihydrogen, which is approximately equal to the positron mass. Additionally, $n$ represents the principal quantum number, while $L$ stands for the orbital angular momentum one.

The energy shift for the $2P_{c-}$ state is given by
\bea
\de \ep_{2P_{c-}}&=&-\sum_{\fb,k}\vev{\pmag^k}_{21}\left(\xi_{\fb,1,c}^{(0B)}\TzBnrf{\fb}{k10}+\xi_{\fb,1,c}^{(1B)}\ToBnrf{\fb}{k10}\right)\nn\\
&&-\sum_{\fb,k}\vev{\pmag^k}_{21}\left(\xi_{\fb,3,c}^{(0B)}\TzBnrf{\fb}{k30}+\xi_{\fb,3,c}^{(1B)}\ToBnrf{\fb}{k30}\right)\nn\\
&&-\fr{1}{\sqrt{5\pi}}\left(\la_1^2-\fr{\la_2^2}{2}\right)\sum_{\fb,k}\Vnrf{\fb}{k20}\vev{\pmag^k}_{21}\nn\\
&& -\fr{1}{\sqrt{4\pi}}\sum_{\fb,k}\Vnrf{\fb}{k00}\vev{\pmag^k}_{21}.
\label{e2Pc}
\eea
Here, $\xi_{\fb,j,c}^{(0B)}$ and $\xi_{\fb,j,c}^{(1B)}$ are dependent on the coefficients $\lambda_1$ and $\lambda_2$. The explicit expressions for the positron case ($\fb=\bar{e}$) are
\bea
\xi_{\bar{e},1,c}^{(0B)}&=&\sqrt{\fr{3}{100\pi}}(3\la_1^2-\sqrt{8}\la_1\la_2 -\la_2^2),\nn\\
\xi_{\bar{e},1,c}^{(1B)}&=&\sqrt{\fr{3}{25\pi}}(\la_1^2+\sqrt{2}\la_1\la_2 -2\la_2^2),\nn\\
\xi_{\bar{e},3,c}^{(0B)}&=&\fr{3}{10\sqrt{7\pi}}(2\la_1^2+\sqrt{8}\la_1\la_2 +\la_2^2),\nn\\
\xi_{\bar{e},3,c}^{(1B)}&=&\sqrt{\fr{6}{175\pi}}(2 \la_1^2+\sqrt{8}\la_1\la_2 +\la_2^2),\nn\\
\label{xice}
\eea
and for the antiproton ($\fb=\bar{p}$)
\bea
\xi_{\bar{p},1,c}^{(0B)}&=&-\sqrt{\fr{3}{100\pi}}(3\la_1^2+\la_2^2),\nn\\
\xi_{\bar{p},1,c}^{(1B)}&=&-\sqrt{\fr{3}{25\pi}}(\la_1^2+2\la_2^2),\nn\\
\xi_{\bar{p},3,c}^{(0B)}&=&-\fr{3}{10\sqrt{7\pi}}(2\la_1^2-\la_2^2),\nn\\
\xi_{\bar{p},3,c}^{(1B)}&=&-\sqrt{\fr{6}{175\pi}}(2\la_1^2-\la_2^2).\nn\\
\label{xicp}
\eea
The expression for the energy shift $\de \ep_{2P_{f-}}$ is obtained by substituting $\lambda_1 \rightarrow \lambda_2$ and $\lambda_2 \rightarrow -\lambda_1$ in \rf{e2Pc}, as can be inferred from the form of the energy states in \rf{2Pstate}. This substitution alters the second-to-last term in \eqref{e2Pc} and transforms the coefficients $\xi_{\fb,j,c}^{(0B)}$ and $\xi_{\fb,j,c}^{(1B)}$ into $\xi_{\fb,j,f}^{(0B)}$ and $\xi_{\fb,j,f}^{(1B)}$, respectively. These coefficients are obtained by applying the aforementioned replacement in \rf{xice} and \rf{xicp}.

Table \ref{xicoeff} provides the numerical values for these coefficients, considering the specific case where $\lambda_1=0.37$ and $\lambda_2=0.93$. The first column lists the corresponding coefficients, with the first two rows showing those for the energy level $2P_{c_{-}}$ and the following two rows for the energy level $2P_{f_{-}}$. The first two columns contain the values relevant to the positron terms. The first column relates to coefficients with $j=1$, while the second column refers to those with $j=3$. The last two columns provide the values relevant to the antiproton case, where the third column corresponds to coefficients with $j=1$ and the fourth column corresponds to those with $j=3$.

\renewcommand\arraystretch{1.5}
\begin{table}
\caption{
Numerical values for the coefficients $\xi_{\fb,j,c}^{(0B)}$, $\xi_{\fb,j,c}^{(1B)}$,  $\xi_{\fb,j,f}^{(0B)}$ and $\xi_{\fb,j,f}^{(1B)}$ when  $\lambda_1=0.37$ and $\lambda_2=0.93$.} 
\setlength{\tabcolsep}{6pt}
\begin{tabular}{ccccc}
\hline
\hline
&  $\fb=\bar{e}$ &  $\fb=\bar{e}$   &  $\fb=\bar{p}$   &  $\fb=\bar{p}$\\
&  $j=1$ &  $j=3$   &  $j=1$   &  $j=3$\\
\hline
$\xi_{\fb,j,c}^{(0B)}$   & $ -0.14 $   &  $ 0.14 $ &  $ -0.12 $ &  $ 0.037$  \\
$\xi_{\fb,j,c}^{(1B)}$   & $ -0.21 $   &  $ 0.22 $ &  $ -0.36 $ &  $ 0.061$  \\
$\xi_{\fb,j,f}^{(0B)}$   & $ 0.33 $   &  $ 0.057 $ &  $  -0.27 $ &  $ -0.10$  \\
$\xi_{\fb,j,f}^{(1B)}$   & $ 0.019 $   &  $ 0.093 $ &  $ -0.22 $ &  $ -0.17$  \\
\hline\hline
\end{tabular}
\label{xicoeff}
\end{table}

It is worth mentioning that the Lorentz-violating contribution to the $2P$ state includes contributions from nonrelativistic coefficients for Lorentz violation with indices $j>1$. However, this is not the case for the commonly targeted transitions in antihydrogen experiments, which involve only $nS$ states, such as the hyperfine transition of the ground state or the $1S$-$2S$ transition. In general, the NR coefficients that can contribute to the $nS$ energy states, as detailed in \cite{kv15}, are the same ones that appear in \rf{e1S}, and they are limited to values of $j$ in the range $j\leq1$. As mentioned in the introduction, one of the main reasons for this work is to use the $1S_{d}-2P_{c-}$ and $1S_{d}-2P_{f-}$ transitions to impose limits on SME coefficients that cannot be studied using the other commonly targeted transitions in antihydrogen.

The Lorentz-violating frequency shifts for the transitions $1S_{d}-2P_{c-}$, denoted as $\de \nu_c$ for the purpose of this work, and $1S_{d}-2P_{f-}$, denoted as $\de \nu_f$, are expressed as
\bea
2\pi \de \nu_c&=&\de \ep_{2P_{c-}}-\de \ep_{1S_d},\nn\\
2\pi \de \nu_f&=&\de \ep_{2P_{f-}}-\de \ep_{1S_d}.
\eea
For instance, the explict expression for $2\pi \de \nu_f$ takes the form
\bea
2\pi \de \nu_f&=&-\sum_{\fb,k}\vev{\pmag^k}_{21}\left(\xi_{\fb,1,f}^{(0B)}\TzBnrf{\fb}{k10}+\xi_{\fb,1,f}^{(1B)}\ToBnrf{\fb}{k10}\right)\nn\\
&&-\sum_{\fb,k}\vev{\pmag^k}_{21}\left(\xi_{\fb,3,f}^{(0B)}\TzBnrf{\fb}{k30}+\xi_{\fb,3,f}^{(1B)}\ToBnrf{\fb}{k30}\right)\nn\\
&&-\fr{1}{\sqrt{12\pi}}\sum_{\fb,k}(\TzBnrf{\fb}{k10}+2\ToBnrf{\fb}{k10})\vev{\pmag^k}_{10}\nn\\
&&-\fr{1}{\sqrt{5\pi}}\left(\la_2^2-\fr{\la_1^2}{2}\right)\sum_{\fb,k}\Vnrf{\fb}{k20}\vev{\pmag^k}_{21}\nn\\
&& -\fr{1}{\sqrt{4\pi}}\sum_{\fb,k}\Vnrf{\fb}{k00}(\vev{\pmag^k}_{21}-\vev{\pmag^k}_{10}).
\label{f2Pf}
\eea

The expression for the Lorentz-violating shift provided above is defined in a laboratory frame where the z-axis is aligned with the magnetic field. In the following section, we will derive the expression for the frequency shift as observed in the Sun-centered frame, which is the reference frame commonly used to report bounds on the coefficients for Lorentz violation.

\section{Lorentz-violating frequency shift in the Sun-centered frame}
\label{sec3}

This section details the derivation of the expression for the frequency shift in the Sun-centered frame. The discussion herein parallels that presented in previous works \cite{gkv14, kv15,kv18}.

The SME coefficients transform as tensor components under observer transformations. Consequently, reporting all constraints on the SME coefficients within the same inertial reference frame is crucial for comparing results across different experiments. The Sun-centered celestial-equatorial frame is commonly used for this role in the literature \cite{sunframe}. In this frame, the coordinates $T$, $X$, $Y$, and $Z$ are defined as follows: the origin of the frame corresponds to the Sun's location during the 2000 vernal equinox; the time $T$ is measured relative to the rest frame of the Sun; the $Z$ axis aligns along the Earth's rotation axis; the $X$ axis points from the Earth to the Sun at $T=0$; and the $Y$ axis completes a right-handed coordinate system.

The boost speed $\beta$ between a laboratory frame on the surface of the Earth and the Sun-centered frame is a small parameter with a value around $10^{-4}$ in natural units. Hence, we can approximate the Lorentz transformation between both frames as a pure rotation and ignore the boost effect. In the Lorentz-violating frequency shift derivation presented in the previous section, we assumed a laboratory frame with its $z$-axis aligned to the applied magnetic field. Therefore, we need to transform the expression for the frequency shift from this laboratory frame to the Sun-centered frame described above.

The nonrelativistic coefficients transform as components of spherical tensors under observer rotations. Consequently, they adhere to simple rotation rules dictated by the rank $j$ of the spherical tensor. It's worth noting that only coefficients with $m=0$ contribute to the frequency shift \rf{f2Pf}. Considering that $m$ specifies the component of the Lorentz-violating spherical tensor, we can interpret this observation as indicating that only the projection of the coefficients in the direction of the applied magnetic field contributes to the frequency shift. In the laboratory frame, this corresponds to the coefficients with $m=0$.

The expression for a generic laboratory frame coefficient ${\K_\f}_{kj0}^{\rm NR}$ with $m=0$ in terms of Sun-centered frame coefficients ${\K_\f}_{kjm}^{\rm NR, Sun}$ is 
\beq
{\K_\f}_{kj0}^{\rm NR} =
\sum_{m=-j}^j e^{i m\om_\oplus \TL}
d^{j}_{0m}(-\chM)
{\K_\f}^{\rm NR,Sun}_{kjm}.
\label{ltos}
\eeq
Here, $\chM$ represents the angle between the applied magnetic field and the Earth's rotation axis that for the particular case of the ALPHA collaboration is $\chM\simeq 70^\circ$ \cite{privALPHA}. The quantities $d^{j}_{mm'}(-\chM)$ denote the small Wigner matrices, as outlined in (136) of Ref. \cite{km09}, evaluated at $-\chM$.

The transformation in \rf{ltos} highlights that the laboratory frame coefficients depend on time and provides the decomposition of this time variation into harmonics of the Earth's sidereal frequency, denoted as $\omega_\oplus \simeq 2\pi/(23\,\text{h}\, 56\,\text{min})$. The local sidereal time $T_L$ is a convenient offset of the time coordinate $T$ that absorbs any potential phase angle in the exponential arguments.

From \rf{ltos} together with \rf{f2Pf}, we can deduce that frequency shift $\de \nu_f$ in the Sun-centered frame, at zeroth-boost order, has the form  
\bea
{2\pi}\de \nu_f&=&   A_{0}^{(f)}+A_{c}^{(f)}\cos{(\om_\oplus \TL)}+A_{s}^{(f)}\sin{(\om_\oplus \TL)}\nn\\
&&+ A_{c2}^{(f)}\cos{(2\om_\oplus \TL)}+A_{s2}^{(f)}\sin{(2\om_\oplus \TL)}\nn\\
&&+ A_{c3}^{(f)}\cos{(3\om_\oplus \TL)}+A_{s3}^{(f)}\sin{(3\om_\oplus \TL)}.
\label{sunshift0}
\eea
Here, the amplitudes $A_{\xi}^{(f)}$ represent linearly independent combinations of NR coefficients. Thus, identifying experimental studies sensitive to each amplitude is crucial for systematically testing CPT and Lorentz symmetry. The most effective method to impose bounds on the coefficients contributing to the amplitudes in front of the harmonics of the sidereal frequency is by constraining the sidereal variation of the resonance frequency, which arises from Earth's rotation in the Sun-centered frame. As the Earth rotates, the orientation of the system—defined as the antihydrogen with the applied magnetic field—changes relative to an inertial reference frame. In the presence of Lorentz violation, the atom's spectrum might depend on the system's overall orientation, and the slow rotation of the system, along with the Earth, results in the sidereal variation observed in \rf{sunshift0}.

Constraining the term $A_{0}^{(f)}$ in \rf{sunshift0} requires a different method because it is not associated with the time variation in the transition frequency, and therefore, sidereal variation studies are insensitive to it. This term includes contributions from coefficients that are isotropic in space but not spacetime and from anisotropic terms in space that remain constant for rotations around Earth's axis.

One approach to constraining the constant shift, previously used in \cite{gkv14,kv15}, is based on the observation that this term can lead to discrepancies between experimental results obtained in Earth-based laboratories and the theoretical predictions of the conventional Lorentz-invariant theory. The argument is that conventional theory does not account for the constant shift in \rf{sunshift0}, and experimental results will be affected by this shift according to our model. Limits on such discrepancies can thus constrain the magnitude of the Lorentz-violating constant shift. In this work, we will use the agreement between theoretical and experimental results for the antihydrogen $1S$-$2P$ transition, as reported by the ALPHA collaboration \cite{1s2p}, to impose a constraint on $A_{0}^{(f)}$.

We will adopt the approach outlined in prior publications, which established limits on SME coefficients by comparing theoretical predictions with experimental findings for the $1S$-$2S$ transition in positronium \cite{kv15} and muonium \cite{gkv14}, as well as for the Lamb shift in muonium \cite{gkv14}. For this purpose, we exclude the time-dependent contribution to the frequency shift, which would typically be investigated by exploring sidereal variations of the resonance frequency. Therefore, we will take the frequency shift as
\bea
{2\pi}\de \nu_f&\simeq&   A_{0}^{(f)}
\label{sunshiftconstf}
\eea
 where the constant term $A_{0}^{(f)}$ is given by
 
\onecolumngrid
\bea
A_{0}^{(f)}&=& -\sum_{\fb,k}\left[ \fr{1}{\sqrt{4\pi}}(\vev{\pmag^k}_{21}-\vev{\pmag^k}_{10})\, \sVnrf{\fb}{k00}+\fr{1+3 \cos{2\vartheta}}{4\sqrt{5\pi}}\left(\la_2^2-\fr{\la_1^2}{2}\right)\vev{\pmag^k}_{21}\,\sVnrf{\fb}{k20}\right]\nn\\
&&-\cos{\vartheta}\sum_{\fb,k}\left[\left(\xi_{\fb,1,f}^{(0B)}\vev{\pmag^k}_{21}+\fr{\vev{\pmag^k}_{10}}{\sqrt{12\pi}}\right)\sTzBnrf{\fb}{k10}+\left(\xi_{\fb,1,f}^{(1B)}\vev{\pmag^k}_{21}+\fr{\vev{\pmag^k}_{10}}{\sqrt{3\pi}}\right)\sToBnrf{\fb}{k10}\right]\nn\\
&&-\fr{1}{8}(3 \cos{\vartheta}+5\cos{3\vartheta} )\sum_{\fb,k}\vev{\pmag^k}_{21}\left(\xi_{\fb,3,f}^{(0B)}\sTzBnrf{\fb}{k30}+\xi_{\fb,3,f}^{(1B)}\sToBnrf{\fb}{k30}\right).
\label{f2Pf}
\eea
\twocolumngrid

Repeating the same arguments for the frequency shift $\de \nu_c$, we obtain that 
\bea
{2\pi}\de \nu_c&\simeq&   A_{0}^{(c)},
\label{sunshiftconstc}
\eea
where $A_{0}^{(c)}$ is obtained by replacing 
\bea
\lambda_1 &\rightarrow &- \lambda_2,\nn\\
\lambda_2 &\rightarrow & \lambda_1 ,\nn\\
\xi_{\bar{e},j,f}^{(0B)} &\rightarrow & \xi_{\bar{e},j,c}^{(0B)} ,\nn\\
 \xi_{\bar{e},j,f}^{(1B)} &\rightarrow & \xi_{\bar{e},j,c}^{1B)} ,
\eea
in the expression for $A_{0}^{(f)}$.  The following section presents the constraints obtained on the constant frequency shifts \rf{sunshiftconstf} and \rf{sunshiftconstc}, based on the agreement between the theoretical and experimental results for the antihydrogen $1S$-$2P$ transition reported by the ALPHA collaboration \cite{1s2p}.

\section{Constraints and discussion}
\label{sec4}

The ALPHA collaboration measured the resonant frequencies for the antihydrogen transitions $1S_d-2P_{c-}$ and $1S_d-2P_{f-}$ in the presence of a 1T magnetic field \cite{1s2p}. They compared their measurements with theoretical predictions from standard atomic theory and found that the difference between the experimental and theoretical values was $(-81 \pm 76)$ MHz for the first transition and $(108 \pm 81)$ MHz for the second at a one-sigma level  \cite{1s2p}. The reported uncertainty comes from the experimental measurements, as the theoretical uncertainties were significantly smaller. Both results are consistent with zero within a two-sigma interval and for that reason, we interpret the difference between the theoretical and experimental values as smaller than 152 MHz for the $1S_d-2P_{c-}$ transition and smaller than 162 MHz for the $1S_d-2P_{f-}$ transition at a two-sigma level.

We aim to constrain the NR coefficients by attributing any difference between the theoretical and experimental values of the transition frequencies to the corresponding constant term in the Lorentz-violating frequency shift. However, we must be cautious with this assumption, as the theoretical prediction for the frequency depends on constants and parameters determined from experiments. As discussed in \cite{kv15}, frequency shifts like the one in \rf{sunshift0} can, in principle, affect these experimental constants by shifting the transition frequencies used to determine their values. Therefore, we should consider the possibility that NR coefficients could influence the theoretical prediction through shifts in the experimental values of these constants and parameters. At the level of the theoretical calculations done by the ALPHA collaboration, which require uncertainties on the order of 1 MHz, this is not a concern, as the impact of the experimental uncertainties in the constants on the theoretical calculations will be smaller than the precision required by the ALPHA collaboration. This allows us to disregard any potential Lorentz-violating shift in the standard theoretical calculation.

As anticipated, we assume that the Lorentz-violating frequency shift accounts for any discrepancy between the experimental and theoretical values of the transition frequencies. Furthermore, since the time-dependent terms in the Lorentz-violating frequency shift \rf{sunshift0} are better studied through sidereal variation analyses, we will disregard their contributions and focus solely on the constant contribution to the frequency shift. Thus, we take the difference between the experimental value $\nu_{c}^{\rm exp}$ and the theoretical value $\nu_{c}^{\rm th}$ for the $1S_d-2P_{f-}$ transition frequency as
\beq
2\pi(\nu_c^{\rm exp}-\nu_c^{\rm th})=A_0^{(c)}.
\eeq
A similar expression holds for the $1S_d-2P_{f-}$ transition. Using our interpretation of an experimental and theoretical agreement to a two-sigma level, we find that
\bea
6.3\times10^{-16}{\rm GeV}&\geq&|A_0^{(c)}|,\nn\\
6.7\times10^{-16}{\rm GeV}&\geq&|A_0^{(f)}|.
\label{bounds}
\eea
Here, we converted from Hz to GeV within a unit system with $\hbar=1$ by using that
\beq
1{\rm Hz}\simeq 6.582\times 10^{-25} {\rm GeV}.
\eeq
 
\renewcommand{\arraystretch}{1.5}
\begin{table*}
\caption{
\label{Cons}
Constraints on the individual NR coefficients based on the bounds \rf{bounds} compared to the corresponding current best constraints } \setlength{\tabcolsep}{5pt} \begin{tabular}{cllllll} \hline \hline
                                  & \multicolumn{2}{c}{$1S_d$-$2P_{c-}$}                                                                                          & \multicolumn{2}{c}{$1S_d$-$2P_{f-}$}            & \multicolumn{2}{c}{Current Bounds} \\ \hline 
Coefficient                                                  & Electron                               & Proton                                  & Electron                              & Proton                                    & Electron                                     & Proton\\ 
${\K_\f}_{kj0}^{\nr, \rm{Sun}}$                    & $({\rm GeV}^{1-k})$           & $({\rm GeV}^{1-k})$              & $({\rm GeV}^{1-k})$           & $({\rm GeV}^{1-k})$                & $({\rm GeV}^{1-k})$                  & $({\rm GeV}^{1-k})$      \\ \hline 
$\sanrf{\f}{200}$                                         &       $2.1\times 10^{-2}$      &       $2.1\times 10^{-2}$       &       $2.3\times 10^{-2}$      &       $2.3\times 10^{-2}$            &       $4.0\times 10^{-9}$ \cite{kv18}      &       $4.0\times 10^{-9}$ \cite{kv18}  \\
$\scnrf{\f}{200}$                                         &       $2.1\times 10^{-2}$      &       $2.1\times 10^{-2}$       &       $2.3\times 10^{-2}$      &       $2.3\times 10^{-2}$          &      $2.1 \times 10^{-5}$    \cite{kv15}\\
$\sanrf{\f}{400}$                                         &       $2.4\times 10^{10}$      &       $2.4\times 10^{10}$      &       $2.5\times 10^{10}$      &       $2.5\times 10^{10}$           &       $5.0\times 10^{2}$  \cite{kv18}     &       $5.0\times 10^{2}$ \cite{kv18}  \\
$\scnrf{\f}{400}$                                         &       $2.4\times 10^{10}$      &       $2.4\times 10^{10}$      &       $2.5\times 10^{10}$      &       $2.5\times 10^{10}$           &       $3.1\times 10^{7}$ \cite{kv15}     \\
$\sanrf{\f}{220}$, $\scnrf{\f}{220}$             &       $7.5\times 10^{-1}$      &      $7.5\times 10^{-1}$        &       $3.0\times 10^{-1}$      &       $3.0\times 10^{-1}$\\
$\sanrf{\f}{420}$, $\scnrf{\f}{420}$             &       $9.3\times 10^{12}$      &       $9.3\times 10^{12}$      &       $3.7\times 10^{12}$       &       $3.7\times 10^{12}$\\
$\sHzBnrf{\f}{010}$, $\sgzBnrf{\f}{010}$    &       $7.8\times 10^{-14}$     &      $4.8\times 10^{-15}$     &       $4.0\times 10^{-15}$     &      $1.9\times 10^{-15}$           &       $7.7\times 10^{-19}$ \cite{no24}    &       $1.2\times 10^{-21}$ \cite{no24}\\
$\sHoBnrf{\f}{010}$, $\sgoBnrf{\f}{010}$   &       $1.7\times 10^{-14}$     &      $4.8\times 10^{-14}$     &       $5.7\times 10^{-15}$      &      $1.9\times 10^{-14}$          &       $3.8\times 10^{-19}$ \cite{no24}      &       $5.8\times 10^{-22}$ \cite{no24}\\
$\sHzBnrf{\f}{210}$, $\sgzBnrf{\f}{210}$    &      $1.0\times 10^{-1}$        &      $1.0\times 10^{-2}$       &      $5.7\times 10^{-2}$        &      $1.5\times 10^{-1}$             &       $5.5\times 10^{-8}$ \cite{no24}       &       $8.4\times 10^{-11}$ \cite{no24}\\
$\sHoBnrf{\f}{210}$, $\sgoBnrf{\f}{210}$   &      $4.9\times 10^{-2}$        &      $5.6\times 10^{-2}$       &      $4.3\times 10^{-2}$        &      $5.2\times 10^{-2}$             &       $2.8\times 10^{-8}$ \cite{no24}        &       $4.2\times 10^{-11}$ \cite{no24}\\
$\sHzBnrf{\f}{410}$, $\sgzBnrf{\f}{410}$    &      $1.2\times 10^{11}$       &      $1.2\times 10^{11}$       &       $1.2\times 10^{11}$       &      $1.3\times 10^{11}$            &       $8.0\times 10^{2}$ \cite{no24}         &       $1.2\times 10^{0}$ \cite{no24}\\
$\sHoBnrf{\f}{410}$, $\sgoBnrf{\f}{410}$   &     $6.0\times 10^{10}$        &      $6.0\times 10^{10}$       &       $6.2\times 10^{10}$       &      $6.4\times 10^{10}$            &       $4.0\times 10^{2}$ \cite{no24}         &       $6.0\times 10^{-1}$ \cite{no24}\\
$\sHzBnrf{\f}{230}$, $\sgzBnrf{\f}{230}$    &      $3.2\times 10^{-1}$        &      $1.2\times 10^{0}$         &      $8.2\times 10^{-1}$        &      $4.6\times 10^{-1}$\\ 
$\sHoBnrf{\f}{230}$, $\sgoBnrf{\f}{230}$   &      $2.0\times 10^{-1}$        &      $7.2\times 10^{-1}$       &      $5.0\times 10^{-1}$         &      $2.8\times 10^{-1}$\\ 
$\sHzBnrf{\f}{430}$, $\sgzBnrf{\f}{430}$    &     $4.0\times 10^{12}$        &      $1.4\times 10^{13}$        &      $1.0\times 10^{13}$        &      $5.7\times 10^{12}$\\
$\sHoBnrf{\f}{430}$, $\sgoBnrf{\f}{430}$   &     $2.5\times 10^{12}$        &      $8.8\times 10^{12}$        &      $6.2\times 10^{12}$        &      $3.5\times 10^{12}$\\
\hline
\hline
\end{tabular}
\end{table*} 

The bounds \rf{bounds} are on a linear combinations of NR coefficients such as the one in \rf{f2Pf}, where $\theta \simeq 70^\circ$, $\lambda_1 = 0.37$, $\lambda_2 = 0.93$, and the $\xi_{\fb,j,f}^{(0B)}$ coefficients are those listed in Table \ref{xicoeff}. It is common in the literature to estimate the maximum individual bound on each coefficient for Lorentz violation by obtaining the bound for one coefficient while assuming all others are zero \cite{tables}. The results of these calculations are in Table \ref{Cons}, where the first column specifies the coefficient for Lorentz violation. Note that the coefficients listed in the table are those with a definite CPT sign that are related to those used in \rf{f2Pf} by \rf{cpt}. Therefore, we will interpret these as bounds on the electron and proton coefficients, rather than on the positron and antiproton coefficients, as is commonly done in the literature \cite{tables}.

The second to seventh columns alternate between the bounds on the size of the electron and proton coefficients. The second and third columns show the bounds obtained from the $1S_d-2P_{c-}$ transition, while the fourth and fifth columns correspond to those obtained from the $1S_d-2P_{f-}$ transition. The last two columns show the current best sensitivity on the corresponding coefficients, with blank entries indicating no previous bounds exist for those coefficients.  

The bounds obtained in this work on previously constrained coefficients were weaker than the best current bounds, which come from comparisons between the hydrogen and antihydrogen $1S$-$2S$ transitions by the ALPHA collaboration \cite{kv18}, from the comparison between theoretical and experimental values of the positronium $1S$-$2S$ transition \cite{kv15}, and from measurements of the Zeeman-hyperfine transitions of the hydrogen ground state using different magnetic field orientations \cite{no24}. These previous measurements have lower absolute uncertainty than the $1S$-$2P$ transitions considered in this work, making them more sensitive to individual NR coefficients. Note that these previous bounds, with the exception of those on the $c$-type coefficients, can be further improved by measuring transitions between $nS$ states in antihydrogen. Specifically, the bounds on the $a$-type coefficients will improve along the measurement of the $1S$-$2S$ transition in antihydrogen, and the ones on the $g$- and $H$-type coefficients could be improved through high-precision measurements of magnetic-sensitive transitions within the $1S$ state in antihydrogen. Despite the advantages of these other transitions over the $1S$-$2P$ transition in antihydrogen, we imposed bounds on 26 previously unconstrained NR coefficients as these other transitions are only sensitive to coefficients with $j \leq 1$, as demonstrated in \cite{kv15}, and are therefore insensitive to 24 of the previously unconstrained coefficients, which can be accessed using the $1S_d-2P_{c-}$ and $1S_d-2P_{f-}$ transitions.

The general rule for hydrogen and antihydrogen is that transitions involving energy states with total electronic angular momentum $J$ and total atomic angular momentum $F$ are insensitive to $\V$-type coefficients with $j > 2J - 1$ and $\T$-type coefficients with $j > 2F - 1$ \cite{kv15}. Therefore, the $1S$-$2S$ transition and the $1S$ splitting are only sensitive to coefficients with $j \leq 1$. However, the situation differs for $1S$-$2P$ transitions, as the $2P_{c-}$ and $2P_{f-}$ energy states receive contributions from angular momentum states with $F = 2$ and $J = 3/2$. The constraints listed in Table I for coefficients with $j > 1$ can be improved not only by refining the measurement of the $1S$-$2P$ transition but also by targeting other transitions that can be measured with lower absolute uncertainty and involve energy states that receive contributions from angular momentum states with $F = 2$ and $J = 3/2$. Furthermore, targeting transitions involving energy states that receive contributions from angular momentum states with $J = 5/2$ would allow for bounds on $\V$-type coefficients with $j = 4$, and if they involve contributions from states with $F = 3$, they would permit bounds on $\T$-type coefficients with $j = 5$. Before moving on, another possibility for improving the bounds or imposing the first bounds on coefficients with $j>1$ is molecular antihydrogen ion spectroscopy \cite{ah2}. As previously argued \cite{kv15}, one key advantage of $\bar{H}^-_2$ spectroscopy is its potential sensitivity to these coefficients.

The reader may have noticed that the smaller numerical bounds in Table \ref{Cons} are for the coefficients with index $k=0$, while the larger ones correspond to those with $k=4$. It is not straightforward to compare these values without making assumptions about the underlying fundamental theory that emerges when the effective field theory breaks down, as coefficients with different values of $k$ have different mass dimensions, given by ${\rm GeV}^{1-k}$. The primary reason for the differences in these bound sizes is the small value of the momentum of the proton and electron within the hydrogen atom when expressed in ${\rm GeV}$ units. Since the $k$ index represents the power of the momentum magnitude associated with the Lorentz-violating operator, higher values of $k$ result in smaller numerical bounds in ${\rm GeV}^{1-k}$ units. In ${\rm eV}^{1-k}$ units, the situation is reversed, with relatively larger numerical values for the momentum and smaller numerical values for the bounds for $k=4$ compared to those for $k=0$ highlighting the issue of comparing bounds on coefficients of different mass dimensions.

Finally, for ordinary matter, a simplified deuterium model and order-of-magnitude calculations have shown that conducting experiments similar to those involving hydrogen but using deuterium could significantly improve the bounds on proton coefficients with $k=2$ and $k=4$ by many orders of magnitude \cite{kv15}. A recently derived model for testing Lorentz and CPT symmetry with deuterium demonstrated that improvements of 5 to 15 orders of magnitude in these bounds are achievable. This improvement is due to the greater momentum of the proton in deuterium, resulting from its motion within the nucleus. The same argument applies to antideuterium, supporting the case for future antideuterium spectroscopy, as it would be far more sensitive to certain violations of CPT and Lorentz symmetry than antihydrogen.

\section{Summary}

In this work, we used the agreement between the measured $1S$-$2P$ antihydrogen transition and its theoretical prediction, as reported by the ALPHA collaboration \cite{1s2p}, to impose the bounds presented in \rf{bounds} on a linear combination of NR coefficients, such as \rf{f2Pf}. This analysis yielded the first-ever constraints on 26 NR coefficients for Lorentz and CPT violation, as shown in Table \ref{Cons}.

We arrived at this result by calculating the Lorentz-violating frequency shift for the targeted transitions in the laboratory frame, as detailed in Sec. \ref{seclab}. In Sec. \ref{sec3}, we derived the constant contribution to the frequency shift in the Sun-centered frame. By attributing any discrepancy between the theoretical and experimental values of the $1S$-$2P$ transition frequency to the constant shift, we obtained the bounds \rf{bounds} in Sec. \ref{sec4}. 

In addition, we discussed the advantages of using the $1S$-$2P$ antihydrogen transition for testing Lorentz and CPT symmetry, compared to the commonly targeted $1S$-$2S$ transition and $1S$ splitting. We highlighted that the $1S$-$2P$ transition offers access to a broader set of coefficients for Lorentz violation due to the higher angular momentum quantum number of the energy states involved in the transition.

\section*{Acknowledgments}

We extend our gratitude to the ALPHA collaboration, particularly to Janko Nauta, for providing essential information about their magnetic trap, which was crucial for this work.


\begin{thebibliography}{99}
\bibitem{ck}
D.\ Colladay and V.A.\ Kosteleck\'y,
Phys.\ Rev.\ D {\bf 55}, 6760 (1997);
Phys.\ Rev.\ D {\bf 58}, 116002 (1998).

\bibitem{ksp}
V.A.\ Kosteleck\'y and S.\ Samuel,
Phys.\ Rev.\ D {\bf 39}, 683 (1989);
V.A.\ Kosteleck\'y and R.\ Potting,
Nucl.\ Phys.\ B {\bf 359}, 545 (1991);
Phys.\ Rev.\ D {\bf 51}, 3923 (1995). 

\bibitem{owg}
O.W.\ Greenberg,
Phys.\ Rev.\ Lett.\ {\bf 89}, 231602 (2002).

\bibitem{anH}
R.\ Bluhm, V.A.\ Kosteleck\'y, and N.\ Russell,
Phys.\ Rev.\ Lett.\ {\bf 82}, 2254 (1999).

\bibitem{pen}
R.\ Bluhm, V.A.\ Kosteleck\'y, and N.\ Russell,
Phys.\ Rev.\ Lett.\ {\bf 79}, 1432 (1997);
R.\ Bluhm, V.A.\ Kosteleck\'y, and N.\ Russell,
Phys.\ Rev.\ D {\bf 57}, 3932 (1998);
R.K.\ Mittleman, I.I.\ Ioannou, 
H.G.\ Dehmelt, and N.\ Russell, 
Phys.\ Rev.\ Lett.\ {\bf 83}, 2116 (1999).

\bibitem{ko04}
V.A./ Kosteleck\'y,
Phys. Rev. D {\bf 69}, 105009 (2004).

\bibitem{grav}
V.A.\ Kosteleck\'y and J.D. Tasson, Phys.\ Rev.\ D {\bf 83},
016013 (2011); Phys.\  Rev.\ Lett.\ {\bf 102}, 010402 (2009).

\bibitem{km09}
V.A.\ Kosteleck\'y and M.\ Mewes,
Phys.\ Rev.\ D {\bf 80}, 015020 (2009).

\bibitem{km12}
V.A.\ Kosteleck\'y and M.\ Mewes,
Phys.\ Rev.\ D {\bf 85}, 096005 (2012);
J.S.\ D\'\i az, V.A.\ Kosteleck\'y, and M.\ Mewes,
Phys.\ Rev.\ D {\bf 89}, 043005 (2014);
V.A.\ Kosteleck\'y and Z.\ Li, 
Phys.\ Rev.\ D {\bf 99}, 056016 (2019).

\bibitem{km13}
V.A.\ Kosteleck\'y and M.\ Mewes,
Phys.\ Rev.\ D {\bf 88}, 096006 (2013).

\bibitem{nonmingrav}
V.A.\ Kosteleck\'y and M.\ Mewes,
Phys.\ Lett.\ B {\bf 757}, 510 (2016); 
Q.G.\ Bailey, V.A.\ Kosteleck\'y, and R.\ Xu,
Phys.\ Rev.\ D {\bf 91}, 022006 (2015);
V.A.\ Kosteleck\'y and J.D.\ Tasson,
Phys.\ Lett.\ B {\bf 749}, 551 (2015).

\bibitem{kl21}
V.A.\ Kosteleck\'y and Z.\ Li, 
Phys.\ Rev.\ D {\bf 103}, 024059 (2021).

\bibitem{dk16}	
Y.\ Ding and M.F.\ Rawnak, 
Phys. Rev. D {\bf 102}, 056009 (2020);
Y.\ Ding and V.A.\ Kosteleck\'y,
Phys.\ Rev.\ D {\bf 94}, 056008 (2016).

\bibitem{kv15}
V.A.\ Kosteleck\'y and A.J.\ Vargas, 
Phys.\ Rev.\ D {\bf 92}, 056002 (2015).

\bibitem{BASE}
M.J.\ Borchert \etal,
Nature {\bf 601}, 53 (2022);
C.\ Smorra \etal,
Nature {\bf 575}, 310 (2019);
S.\ Ulmer \etal,
Nature {\bf 524},196 (2015).




\bibitem{ah17}
M.\ Ahmadi \etal, 
Nature {\bf 548}, 71 (2017).

\bibitem{ah18}
M.\ Ahmadi \etal, 
Nature {\bf 557}, 71 (2018);
Nature {\bf 541}, 506 (2017).

\bibitem{ah18a}
M.\ Ahmadi \etal, 
Nature  {\bf 561}, 211 (2018).

\bibitem{1s2p}
M.\ Ahmadi \etal, 
Nature {\bf 578}, 375 (2020).

\bibitem{an23}
E K.\ Anderson \etal, 
Nature {\bf 621}, 716 (2023).

\bibitem{gbar}
P.\ P\'erez \etal, 
Hyperfine Interact. {\bf 233}, 21 (2015).

\bibitem{aegis}
A.\ Kellerbauer \etal.,
Nucl. Instr. Meth. B 266, 351-356 (2008).

\bibitem{ASACUSA}
E.\ Widmann \etal,
Hyperfine Interact. {\bf 215}, 1 (2013).

\bibitem{kv18}
V.A.\ Kosteleck\'y and A.J.\ Vargas,
Phys.\ Rev.\ D {\bf 98}, 036003 (2018).

\bibitem{gkv14}
A.H.\ Gomes, V.A. Kosteleck\'y, and A.J.\ Vargas,
Phys.\ Rev.\ D {\bf 90}, 076009 (2014).

\bibitem{sunframe}
V.A.\ Kosteleck\'y and M.\ Mewes,
Phys.\ Rev.\ D {\bf 66}, 056005 (2002).

\bibitem{privALPHA}
ALPHA collaboration (private communication).

\bibitem{tables}
V.A.\ Kosteleck\'y and N.\ Russell,
Rev.\ Mod.\ Phys.\ {\bf 83}, 11 (2011);
2023 edition arXiv:0801.0287v16.

\bibitem{no24}
L.\ Nowak \etal, Phys.\ Lett.\ B {\bf 854}, 139012 (2024).

\bibitem{ah2}
M.R.\ Schenkel, S.\ Alighanbari, and S.\ Schiller, 
Nat. Phys. {\bf 20}, 283 (2024).
M.C.\ Zammit \etal,
Phys.\ Rev.\ A {\bf 100}, 042709 (2019);
E.G. Myers,
Phys.Rev. A {\bf 98}, 010101 (2018).

\bibitem{va24}
A.J.\ Vargas, 
Phys.\ Rev.\ D {\bf 109}, 055001 (2024).

\end{thebibliography}
\end{document}